\documentclass[reprint,NumberedRefs]{JASAnew}
\usepackage{multirow}
\usepackage{graphicx}
\usepackage{booktabs}
\usepackage{setspace}

\begin{document}

\title[JASA/Deep learning for source localization]{Deep-learning source localization using multi-frequency magnitude-only data}
\author{Haiqiang Niu}
\email{Electronic mail: nhq@mail.ioa.ac.cn}
\affiliation{State Key Laboratory of Acoustics, Institute of Acoustics, Chinese Academy of Sciences, Beijing, 100190, People's Republic of China}

\author{Zaixiao Gong}
\affiliation{State Key Laboratory of Acoustics, Institute of Acoustics, Chinese Academy of Sciences, Beijing, 100190, People's Republic of China}

\author{Emma Ozanich}
\affiliation{Scripps Institution of Oceanography, University of California San Diego, La Jolla, California 92093-0238, USA}

\author{Peter Gerstoft}
\affiliation{Scripps Institution of Oceanography, University of California San Diego, La Jolla, California 92093-0238, USA}

\author{Haibin Wang}
\affiliation{State Key Laboratory of Acoustics, Institute of Acoustics, Chinese Academy of Sciences, Beijing, 100190, People's Republic of China}

\author{Zhenglin Li}
\affiliation{State Key Laboratory of Acoustics, Institute of Acoustics, Chinese Academy of Sciences, Beijing, 100190, People's Republic of China}

 

\preprint{Author, JASA}		

\date{\today} 

\begin{abstract}
A deep learning approach based on big data is proposed to locate broadband acoustic sources using a single hydrophone in ocean waveguides with uncertain bottom parameters. Several 50-layer residual neural networks, trained on a huge number of sound field replicas generated by an acoustic propagation model, are used to handle the bottom uncertainty in source localization. A two-step training strategy is presented to improve the training of the deep models. First, the range is discretized in a coarse (5 km) grid. Subsequently, the source range within the selected interval and source depth are discretized on a finer (0.1 km and 2 m) grid. The deep learning methods were demonstrated for simulated magnitude-only multi-frequency data in uncertain environments. Experimental data from the China Yellow Sea also validated the approach.
\end{abstract}


\maketitle

\section{\label{sec:1} Introduction}

Recently, ocean acoustic source localization was obtained by machine learning,\cite{Niu1,Niu2} which achieved a lower range estimation error compared with the conventional matched-field processing (MFP).\cite{Bucker,Westwood,Baggeroer,Michalopoulou} They demonstrated that machine learning trained on observed data sets performed well in ship range localization. This method is particularly advantageous when historical data are available to train machine learning models. Even though more data could be collected by using the Automatic Identification System (AIS), there are still not sufficient databases available for some applications. In addition, collecting acoustic data for every source position (i.e. range and depth) in a large ocean area is impractical. To our knowledge, all localization using machine learning are based on small data sets.\cite{Niu1,Niu2,Steinberg,Ozard,Michalopoulou2,Lefort,Wang,Huang,Liu} Without sufficient training data, the machine learning models trained on small data sets are limited to specific environments. The environmental variation (e.g. bottom parameters) degrades the localization performance.

A challenging task in real world is to locate a source in uncertain ocean environments. For MFP,\cite{Bucker,Westwood,Baggeroer,Michalopoulou,Gemba1,Gemba2} environmental mismatch can significantly affect the localization. Thus the geoacoustic parameters are often included as unknown parameters to account for environmental uncertainty.\cite{Gerstoft1,Gingras,Collins,Dosso1,Baer,Richardson,Dosso2,Dosso3} To solve the source positions, focalization\cite{Gerstoft1,Gingras,Collins,Dosso1,Baer} involves maximizing the posterior probability density over all parameters to seek the globally optimal solution, while marginalization\cite{Richardson,Dosso2,Dosso3} integrates the posterior probability density over environmental parameters to obtain the joint marginal probability distributions for source parameters. The advantage of marginalization is that the joint marginal distribution provides a quantitative measure of localization uncertainty.

Source localization or geoacoustic inversion using a single hydrophone is another challenge. Several time-domain Clay-like estimators\cite{Frazer} were introduced and tested on simulated data for single-hydrophone localization. Another approach is to maximize a mean least square criteria for the distance between two subspaces spanned by the delayed source signal paths.\cite{Jesus} In studies,\cite{Siderius,Hermand,Gac} model-based matched filters were used for broadband coherently processing in the single-hydrophone geoacoustic inversion. An alternative approach\cite{Spain,Cockrell,Rakotonarivo} is using interference patterns, or waveguide invariant, from acoustic spectrograms of one sensor. Generally, only the source range is determined by this method.

In the MFP approaches,\cite{Gerstoft1,Gingras,Collins,Dosso1,Baer,Richardson,Dosso2,Dosso3,Siderius,Hermand,Gac} a large number of field replicas generated by acoustic propagation models were used to reflect the environmental uncertainty. Similarly, in this study, we solve source localization using one sensor by exploiting a large number of replicas, as in MFP, for deep learning,\cite{Deep1,Deep2,Deep3} a state-of-the-art method in machine learning. Deep learning models have been shown to outperform shallow machine learning in image processing,\cite{Krizhevsky} speech recognition\cite{Hinton} and natural language processing.\cite{Collobert} Deep neural networks (DNNs) have more parameters and take advantage of big data. In our study, big data are created by generating a huge number of field replicas using KRAKEN.\cite{Porter} Specifically, the contributions of our work are:

(1) A set of 50-layer residual neural networks,\cite{He} known as ResNet50, are trained separately to determine the source range and depth.

(2) The magnitudes of multi-frequency data from a single sensor are used for localization, showing the remarkable capability of DNNs. Note that the localization methods\cite{Spain,Cockrell,Rakotonarivo} based on waveguide invariant also use just the magnitude pressure.

(3) A two-step training strategy is proposed to alleviate the training difficulty for the wide-region source localization (see Sec.~\ref{subsec:2:3} for details).

(4) Big data (tens of millions of training samples, see Sec.~\ref{subsec:3:2}) are used as the training set. The resulting DNNs are able to locate the source in various environments.

DNN depth is of crucial importance in neural network architectures, but deeper networks are more difficult to train due to the vanishing gradient problem. The ResNet\cite{He} allows for very deep structures by introducing identity shortcut connections and learning the residual functions instead of the original ones. The bottleneck architecture in Sec.~\ref{subsec:2:2} is one of the designs to overcome training saturation. Here, the 50-layer residual neural networks are used as the classifiers to solve the source localization problem. 

The paper is organized as follows: Section~\ref{sec:2} presents the data preprocessing and localization algorithm. Simulation and experimental results are given in Secs.~\ref{sec:3} and \ref{sec:4}, demonstrating the performance of DNNs. In Sec.~\ref{sec:5}, computation time, extension to low-SNR and multi-source cases, and limitations are briefly discussed. The summary is given in Sec.~\ref{sec:6}.

\section{\label{sec:2} Localization based on deep learning}
Source localization is solved by DNN (ResNet50) using broadband data from one receiver. The data preprocessing is given in Sec.~\ref{subsec:2:1}. The structure of the residual neural network is described in Sec.~\ref{subsec:2:2}. The localization algorithm and metric are given in Secs.~\ref{subsec:2:3} and \ref{subsec:2:4}.

\subsection{\label{subsec:2:1} Input data preprocessing}
The complex pressure $\mathbf{p}=[p_1,...,p_f,...,p_F]$ at $F$ frequencies is obtained by taking the discrete Fourier transform of the raw pressure data at the sensor. The sound pressure $p_f$ at frequency $f$ is modeled as
\begin{equation} \label{sound pressure}
p_f=S(f)g(f,\mathbf{r})+\epsilon,
\end{equation}
where $S(f)$ is the source term, $g(f,\mathbf{r})$ is the Green's function, and $\epsilon$ is noise. The magnitude of the pressure at $F$ frequencies is written as the vector
\begin{equation} \label{magnitude1}
\mathbf{q}=\left[~|p_1|,...,|p_f|,...,|p_F|~ \right].
\end{equation}

To facilitate training of DNNs, Eq.~(\ref{magnitude1}) is normalized according to
\begin{equation} \label{normalized_amp}
\tilde{\mathbf{q}}=\frac{\mathbf{q}-\min(\mathbf{q})}{\max\left[\mathbf{q}-\min(\mathbf{q}) \right]},
\end{equation}
where $\min$($\cdot$) and $\max$($\cdot$) represent the minimum and maximum of the vector. Thus the value of each element in $\tilde{\mathbf{q}}$ is within the interval $[0,1]$. 

As indicated in Eq.~(\ref{sound pressure}), the source term $|S(f)|$ may vary for each source, resulting in feature differences between the training and test data. To reduce the effect of source magnitudes, a piecewise normalization method is used for the case of slowly varying source magnitude in frequency domain. The vector $\tilde{\mathbf{q}}$ is divided into several segments: \{$\tilde{\mathbf{q}}(1:n_f)$, $\tilde{\mathbf{q}}(n_f+1:2n_f), \dots$\} with the length of each segment $n_f$. Each segment is normalized to its maximum value according to
\begin{equation} \label{normalized_segment}
\left \{
	\begin{array}{lr}
		\hat{\mathbf{q}}(1:n_f)=\frac{\tilde{\mathbf{q}}(1:n_f)}{\max\left[\tilde{\mathbf{q}}(1:n_f) \right]}, & \\
		\hat{\mathbf{q}}(n_f+1:2n_f)=\frac{\tilde{\mathbf{q}}(n_f+1:2n_f)}{\max\left[\tilde{\mathbf{q}}(n_f+1:2n_f) \right]},&\\
		 \vdots \\
	\end{array}
\right .
\end{equation}

The vector $\hat{\mathbf{q}}$ is a feature vector for a single input sample of the DNN. Denoting the number of training samples as $M$, the dimension of the training set is $M \times F$. $n_f$ is selected empirically as 20 (based on numerical experiments on the validation set, choice of $n_f$ from [10, 25] gave similar performance). Note that the magnitude in Eq.~(\ref{magnitude1}) are equivalent to the Sample Covariance Matrix (SCM) when there is only one sensor. The use of magnitude Eq.~(\ref{magnitude1}) and normalization Eq.~(\ref{normalized_segment}) is to reduce the input difference between training and test data. 

Thus, the localization is based on the magnitude of frequency-domain signal from a single sensor. The piecewise normalization is only valid for the slowly varying source magnitude which is the scenario we consider in this paper.

\subsection{\label{subsec:2:2} Residual neural network}
\begin{figure}[h]
	\fig{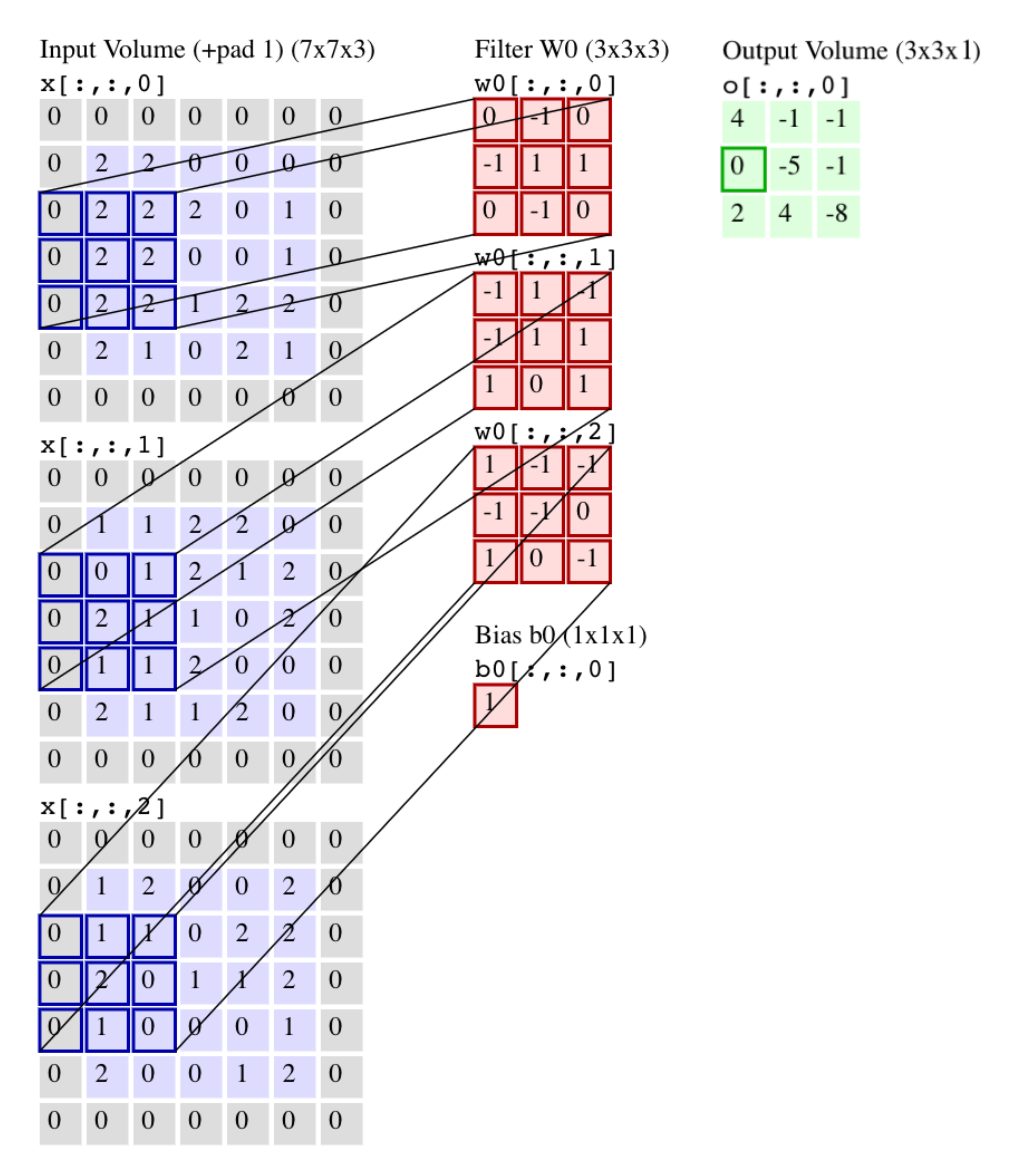}{8.5cm}{}
	\caption{(Color online) Example\cite{cs231n} of a single convolutional layer for $5\times5\times3$ inputs padded with $P = 1$ and one $3\times3\times3$ filter.}
	\label{fig1}
\end{figure}

The residual neural network, ResNet,\cite{He} is a form of convolutional network that uses layers of filters to learn spatial features from the input data. An example\cite{cs231n} of a single convolution layer is shown in Fig.~\ref{fig1}. The filters are typically much smaller than the inputs, allowing them to capture small-scale features in the input. Compared with the fully connected feed-forward neural networks, the number of unknown parameters in the convolution layer is reduced significantly by weight sharing. The $7\times7\times3$ input volume and $3\times3\times3$ filter were sliced along the third dimension to visualize.

In Fig.~\ref{fig1}, input volumes of size $5\times 5 \times 3$ ($W\times H\times D$) have been padded around the perimeter using $P=1$ and sliced for visualization. The last dimension $D$ represents the number of channels in image processing. The output values are the sum of the pointwise multiplication for one filter with $3\times 3 \times 3$ ($F_1\times F_2\times D$) of the input, plus a bias of 1. A slide interval, or stride, of $K=2$ is used in the example. In general, the output dimensions for width and height are
\begin{equation} \label{output_width}
(W-F_1+2P)/K+1,
\end{equation}
and
\begin{equation} \label{output_height}
(H-F_2+2P)/K+1.
\end{equation}
Thus, in Fig.~\ref{fig1}, the output volume has dimensions $3 \times 3 \times 1 ((5-3+2)/2+1=3)$, where the third dimension of 1 indicates a single filter (the size of last dimension is the number of filters). To calculate the boxed cell in output volume, for example, we sum the pointwise multiplication results between the input volume and the filter: $(-2+2+2-2)+(-1+2+1+1)+(-1-1-2)+1 = 0$.

Stacking multiple convolutional layers to form a deep convolutional network improves model inference, but it also causes the gradients to grow exponentially large or small during training, leading to the so-called vanishing gradient problem.\cite{He}

\begin{figure}[h]
	
	\fig{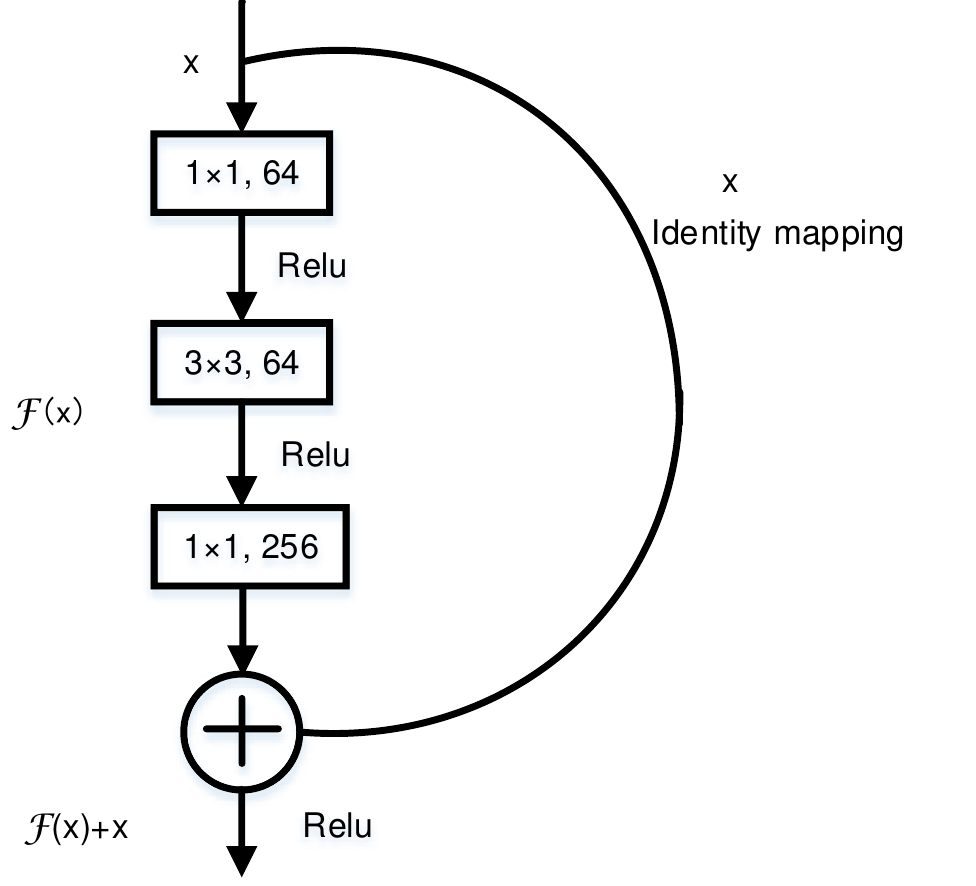}{7.2cm}{}
	\caption{The bottleneck block.}
	\label{fig2}
\end{figure}

The bottleneck architecture (Fig.~\ref{fig2}) of ResNet\cite{He} was introduced to overcome training saturation caused by vanishing gradient. Unlike standard convolutional networks, ResNet models the residual function $\mathcal{F}(\rm x)$, i.e. the difference between a desired output mapping $\mathcal{H}(\rm x)$ and the input $\rm x$: $\mathcal{H}(\rm x)=\mathcal{F}(\rm x)+\rm x$. To achieve this, identity mappings are added to convolutional outputs using a shortcut connection.

Figure~\ref{fig2} shows a block of three convolutional layers with $1\times1$, $3\times3$, and $1\times1$ convolutions. The corresponding numbers of filters are 64, 64, and 256. This architecture first reduces, then enlarges the output dimensions, creating a bottleneck. Common ResNet models with this design use 50, 101, or 152 convolutional layers, made by stacking the three-layer structure in Fig.~\ref{fig2}.

Considering the size of data sets for our localization problem, 50-layer ResNets are used. Our inputs are one-dimensional (see Eq.~(\ref{normalized_amp})), and the 2-D convolutions in ResNets\cite{He} are replaced by 1-D convolutions. The maxpooling operation is also removed in our design due to the small input dimension. The architecture of ResNet50 used here is shown in Table~\ref{tab:table1}. As in the original ResNet,\cite{He} the softmax classifier is used in the last layer of the model and ReLU (Rectified Linear Unit) is used as the activation function:
\begin{equation} \label{sofmax_func}
F_\mathrm{softmax} = \frac{\exp(a_k)}{\sum_{j=1}^{K}\exp(a_j)},
\end{equation}
\begin{equation} \label{ReLU_func}
F_\mathrm{ReLU}(x) = \max(0,x).
\end{equation}

\begin{table}[htb]
	\caption{\label{tab:table1} ResNet50 architecture for source localization.}
	
	\begin{ruledtabular}
		\begin{tabular}{cc}
		   Layer name & [kernel size, filters] $\times$ No. of blocks \\
			\midrule
			conv1 & [7 $\times$ 1, 64] $\times$ 1 \\  
			\midrule
			conv2\_x & 
			$\begin{bmatrix}
			1 \times 1, & 64 \\
			3 \times 1, & 64 \\
			1 \times 1, & 256 \\
			\end{bmatrix} \times 3$ \\
			\midrule
			conv3\_x & 
			$\begin{bmatrix}
			1 \times 1, & 128 \\
			3 \times 1, & 128 \\
			1 \times 1, & 512 \\
			\end{bmatrix} \times 4$ \\
			\midrule
			conv4\_x & 
			$\begin{bmatrix}
			1 \times 1, & 256 \\
			3 \times 1, & 256 \\
			1 \times 1, & 1024 \\
			\end{bmatrix} \times 6$ \\
			\midrule 
			conv5\_x & 
			$\begin{bmatrix}
			1 \times 1, & 512 \\
			3 \times 1, & 512 \\
			1 \times 1, & 2048 \\
			\end{bmatrix} \times 3$ \\
			\midrule
			\multicolumn{2}{c}{average pooling, fully-connected layer with softmax } \\
		\end{tabular}
	\end{ruledtabular}
	
\end{table}
\subsection{\label{subsec:2:3} Localization algorithm}
Training deep learning models becomes harder as the searching space grows. In source localization, the possible ranges may vary in a large scale (e.g. 1--20 km), which increases the training difficulty. 

\begin{figure}[h]	
	\fig{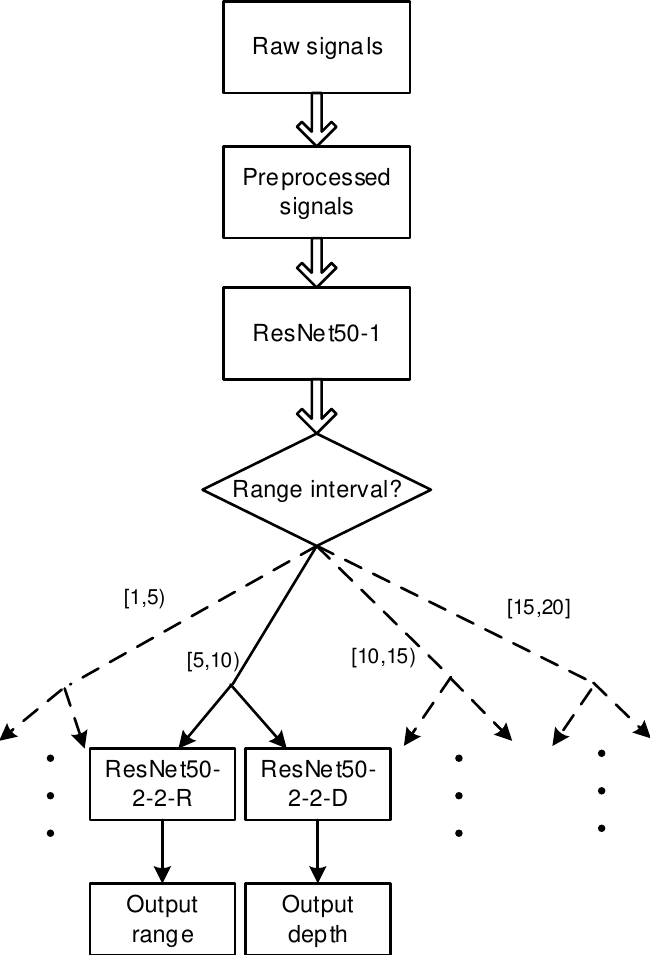}{6.5cm}{}
	\caption{Localization algorithm.}
	\label{fig3}
\end{figure}

To address this problem, a two-step training strategy is used. In the first step, the range is discretized into coarse grids with low range resolution. The range interval 1--20 km is discretized every 5 km to 4 classes: [1,5), [5,10), [10,15), [15,20]. Every input signal is classified by the deep learning model, ResNet50-1, into one of these four range intervals. In the second step, each of these four range intervals is discretized into fine grids. For each range interval, the source range and depth are estimated using ResNet50 classifiers, one for each parameter (Fig.~\ref{fig3}). Overall, one ResNet50 model is trained in step one and eight ResNet50 models are trained in step two. We denote the second-step models as ResNet50-2-x-R and ResNet50-2-x-D, where x indicates which range interval the models are trained under.

As the source localization is treated as a classification problem, the source range and depth are discretized and mapped to a set of binary vectors.\cite{Niu1,Niu2} These binary vectors are used as the labels of the deep learning models. Note that the range labels are designed differently to adapt to the classification tasks in the two-step processing (see Fig.~\ref{fig3} and Sec.~\ref{subsec:3:2}).
\subsection{\label{subsec:2:4} Performance metric}
According to the task difference between the two steps in Sec.~\ref{subsec:2:3}, two metrics quantify the performance.

For the first step, the classification accuracy is used,
\begin{equation} \label{accuracy}
E_{\rm{Accu}}=\frac{N_c}{N}\times 100\%.
\end{equation}  
$N_c$ represents the number of samples classified to the correct range interval and $N$ is the total number of samples. 

For the second step, the mean absolute error (MAE) is used to evaluate the prediction performance
\begin{equation} \label{MAE}
E_{\rm{MAE}}=\frac{1}{N}\sum_{i=1}^{N}\left|Rp_i-Rg_i\right|,
\end{equation}
where $Rp_i$ and $Rg_i$ are the predicted and the ground truth parameters, for either range or depth.

\section{Simulations \label{sec:3}}

\subsection{\label{subsec:3:1} Environmental model and data sets generation}
\begin{figure}[htb]	
	\fig{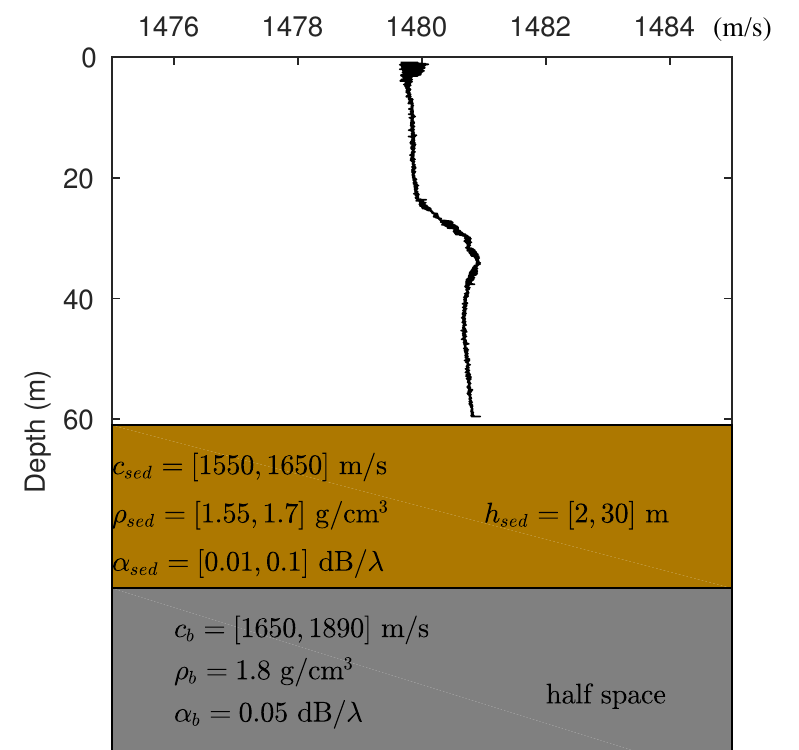}{8.0cm}{}
	\caption{(Color online) Environment for simulation. The sound speed profile of water column was measured in China Yellow Sea.}
	\label{fig4}
\end{figure}

The ocean environment is simulated by a typical shallow water waveguide consisting of a water column with a sound speed profile measured in an at-sea experiment in winter (see Fig.~\ref{fig4}), a sediment layer and a fluid halfspace basement (no shear effects were considered). The source frequencies are 100--200 Hz with the increment 1 Hz. The source and environmental parameters used in simulations are shown in Table~\ref{tab:table2}. The single receiver is 0.2 m above the bottom.
\begin{table}[htb]
	\begin{spacing}{0.8}
	\caption{\label{tab:table2}Environmental parameters used for training and validation sets generation.}
	
		\begin{tabular}{ccccc}
			\hline\hline
			\multicolumn{5}{c}{Training data set }   \\
			\hline
			\multirow{2}*{Parameters}	& \multirow{2}{*}{Units} & Lower & Upper & No. of \\
			&  & bound & bound & discrete values \\
			\hline
			source range & $\rm{km}$ & 1.0 & 20.0 & 191  \\
			source depth & $\rm{m}$  & 2.0 & 60.0 & 30   \\
			water depth  & $\rm{m}$  & 68.0 & 73.0 & 11   \\
			\underline{sediment parameters} & & & & \\
			sediment thickness & $\rm{m}$ & 2.0 & 30.0 & 8 \\
			P-wave speed & $\rm{m/s}$ & 1550 & 1650 & 11 \\
			density & $\rm{g/cm^3}$ & 1.55 & 1.7 & 4 \\
			P-wave attenuation & $\rm{dB/\lambda}$ & 0.01 & 0.1 & 4 \\
			\underline{basement parameters} & & & & \\
			P-wave speed & $\rm{m/s}$ & 1650 & 1890 & 9 \\
			density & $\rm{g/cm^3}$ & 1.8 & 1.8 & 1 \\
			P-wave attenuation & $\rm{dB/\lambda}$ & 0.05 & 0.05 & 1 \\
			\\
			\multicolumn{5}{c}{Validation data set }   \\
			\hline
			\multirow{2}*{Parameters}	& \multirow{2}{*}{Units} & Lower & Upper & No. of \\
			&  & bound & bound & discrete values \\
			\hline
			source range & $\rm{km}$ & 1.0 & 20.0 & 191  \\
			source depth & $\rm{m}$  & 2.0 & 60.0 & 30   \\
			\hline
		    \multicolumn{3}{c}{Environment}  & \\
			water depth  & $\rm{m}$  & 69.0 &  &    \\
			\underline{sediment parameters} & & & & \\
			sediment thickness & $\rm{m}$ & 15.0 &  &  \\
			P-wave speed & $\rm{m/s}$ & 1638 &  &  \\
			density & $\rm{g/cm^3}$ & 1.68 &  & \\
			P-wave attenuation & $\rm{dB/\lambda}$ & 0.05 &  &  \\
			\underline{basement parameters} & & & & \\
			P-wave speed & $\rm{m/s}$ & 1811 &  &  \\
			density & $\rm{g/cm^3}$ & 1.8 &  &  \\
			P-wave attenuation & $\rm{dB/\lambda}$ & 0.05 &  & 		\\
		\hline\hline  
		\end{tabular}
      \end{spacing}
	
\end{table}

Three kinds of data sets including training, validation and test sets are generated by KRAKEN. The training set is used to train the ResNet50 models, the performance on validation set is used to determine the best model parameters, while the test set is used to examine the generalization capability of the trained models. The parameters used for training and validation sets generation are shown in Table~\ref{tab:table2}. The environmental parameters are sampled uniformly from the given intervals. Therefore, the resolution is 0.1 km in range and 2 m in depth. Based on the parameter sensitivity analysis in Sec.~\ref{subsec:3:2}, the P-wave attenuation coefficient for the basement is 0.05 $\rm dB/\lambda$ to reduce the problem complexity. The density for the basement is also fixed. From Table~\ref{tab:table2}, there are 139392 ($11\times 8 \times 11 \times 4 \times 4 \times 9$) possible environments. The generated training data at different source positions for these environments are fed to the ResNet50 models in Sec.~\ref{subsec:3:3}.

\subsection{\label{subsec:3:2} Sensitivity analysis}
The physics using Eq.~(\ref{normalized_segment}) as the input feature is the interference pattern between modes. The waveguide invariant, related to this interference pattern, has been investigated for source localization.\cite{Spain,Cockrell,Rakotonarivo} Physically, the interference structure depends on the source locations and waveguide parameters with different sensitivity. The mean squared error (MSE) is used to evaluate the parameter sensitivity:
\begin{equation} \label{MSE}
E_{\rm{MSE}}=\frac{1}{F}\sum_{f=1}^{F}\left|\hat{\mathbf{q}}_{b}(f)-\hat{\mathbf{q}}_{v}(f)\right|^2,
\end{equation}
where $\hat{\mathbf{q}}_{b}(f)$ represents the input feature at the $f$th frequency from Eq.~(\ref{normalized_segment}) for the baseline waveguide parameters, and $\hat{\mathbf{q}}_{v}(f)$ corresponds the input feature by varying one parameter while keeping the others fixed. The baseline parameters used for sensitivity analysis are set to \{source range 12 km, source depth 28 m, water depth 70 m, sediment thickness 12 m, sediment sound speed 1620 m/s, sediment density 1.6 g/cm$^3$, sediment attenuation 0.02 dB/$\lambda$, basement speed 1800 m/s, basement density 1.8 g/cm$^3$, basement attenuation 0.05 dB/$\lambda$\}. The input feature is more sensitive to the parameters with larger MSE.

\begin{figure*}[tb]		
	\fig{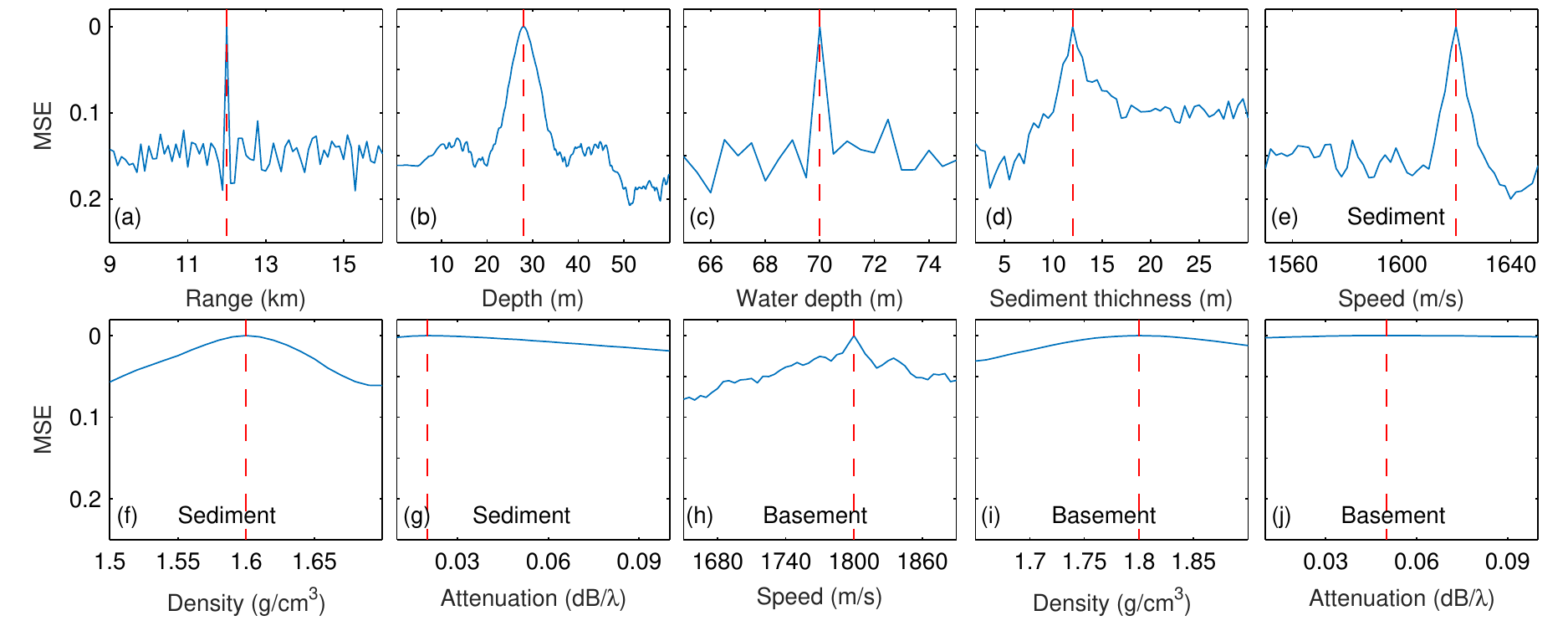}{18.0cm}{}
	\caption{(Color online) Parameter sensitivity analysis for (a) source range, (b) source depth, (c) water depth, (d) sediment thickness, (e) sediment sound speed, (f) sediment density, (g) sediment attenuation, (h) basement sound speed, (i) basement density, and (j) basement attenuation. The dashed lines denote the baseline parameters.}
	\label{fig5}
\end{figure*}

The MSE for 10 parameters is given in Fig.~\ref{fig5}. The MSE shows that the source range and depth, water depth, sediment thickness and sound speed are more sensitive than the other parameters, thus dominating the input features. The MSE for basement density and attenuation vary little across the intervals. Based on the sensitivity analysis, the training set is generated with more possibilities on the sensitive parameters as shown in Table \ref{tab:table2}.  

\subsection{\label{subsec:3:3} ResNet50 model parameters}
For the input sample, ResNet50 determines the range interval in the first step and four pairs of ResNet50 models (ResNet50-2-x-R and ResNet50-2-x-D with x=1,2,3,4) determine the output range and depth across a finer grid in the second step. In these two steps, the source ranges and corresponding labels are different, as shown in Table~\ref{tab:table3}. To generate the training data, the range and depth are discretized every 0.1 km and 2 m respectively. We use range overlaps for the ResNet models in the second step to improve the robustness of localization. The
resulting training data have between $2\times10^8$ and $8\times10^8$ samples, depending on the ResNet50 model and its parameter search space. The models are implemented using Keras\cite{Chollet} with Tensorflow\cite{Tensorflow} as backend.
\begin{table}[tb]
	\caption{\label{tab:table3}Parameters of ResNet50 models.}
	
	\begin{ruledtabular}
		\begin{tabular}{cccc}
			\multirow{2}*{Model} & Range interval  & No. of & No. of \\
			& (km) & range classes & depth classes \\
			\hline
			ResNet50-1 & [1, 20] & 4 & --   \\
			\hline
			ResNet50-2-1-R & [1, 6]  &  51 & --    \\
			ResNet50-2-1-D  & [1, 6]  & -- & 30   \\
			ResNet50-2-2-R & [4, 11]  & 71 & --    \\
			ResNet50-2-2-D  & [4, 11]  & --& 30   \\
			ResNet50-2-3-R & [9, 16]  & 71 & --    \\
			ResNet50-2-3-D  & [9, 16]  & --& 30   \\	
			ResNet50-2-4-R & [14, 20] & 61 & --    \\
			ResNet50-2-4-D  & [14, 20] & --& 30   \\	  	
		\end{tabular}
	\end{ruledtabular}	
\end{table}

\subsection{\label{subsec:3:4} Results}

\subsubsection{\label{subsec:3:4:1} Training and validation errors}
\begin{figure}[tb]	
	\fig{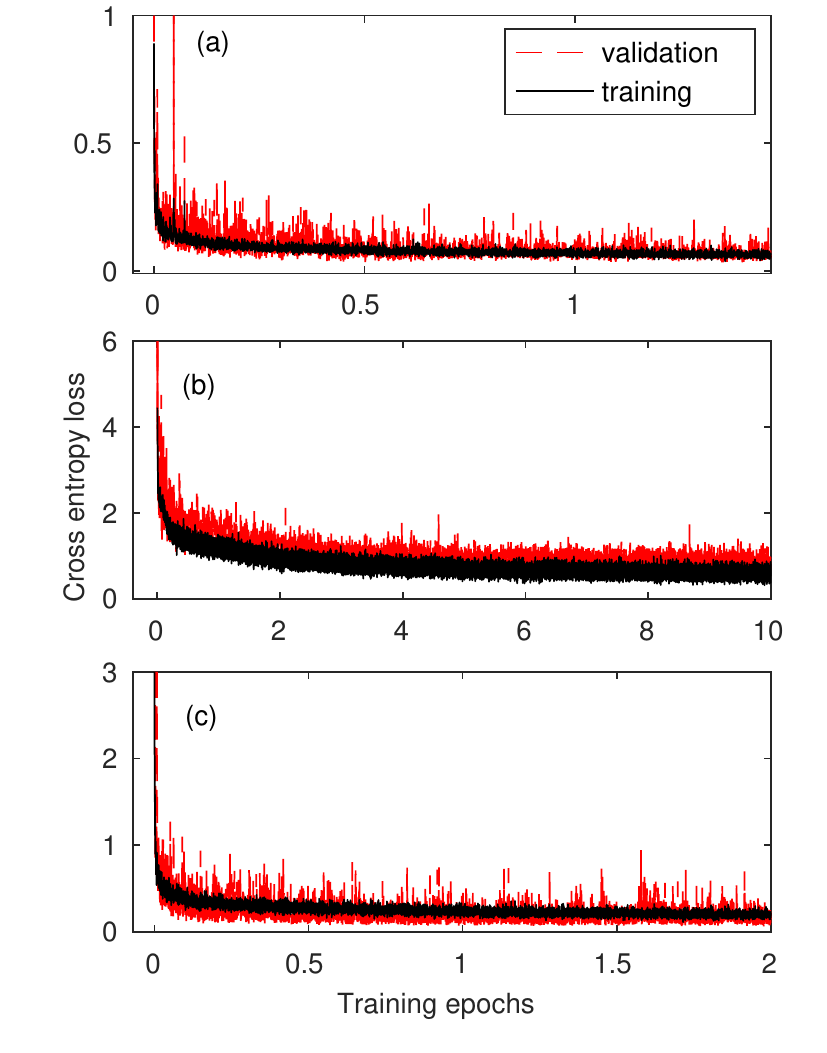}{7.5cm}{}
	\caption{(Color online) Training and validation loss versus training epochs for (a) ResNet50-1 with 1.5 epochs, (b) ResNet50-2-2-R with 10 epochs, and (c) ResNet50-2-2-D with 2 epochs.}
	\label{fig6}
\end{figure}

A validation data set is generated using the randomly selected parameters, shown in Table~\ref{tab:table2}, to examine the validation loss during the training process. In the first training step, the ResNet50-1 model is trained with 1.5 epochs (One epoch consists of one full training cycle on the training set). The model is updated $N/N_b$ times for one epoch, where $N_b=384$ is the batch size for training and $N \in [2\times10^8,8\times10^8] $ is the number of training samples. For the multi-classification problem, the averaged cross entropy\cite{Niu1,Niu2} is used as the loss function:
\begin{equation} \label{CE}
E_{\rm{CE}}=-\frac{1}{N_b}\sum_{n=1}^{N_b}\sum_{k=1}^{K}t_{nk}\ln{y_{nk}},
\end{equation}
where $K$ is the number of classes for the model output, $t_{nk}$ is the $k$th element of label vector for the $n$th sample, and $y_{nk}$ represents the $k$th output. The training and validation loss for the ResNet50-1 is shown in Fig.~\ref{fig6}(a). In the second step, there are two ResNet50 models for each range interval. The training and validation losses of ResNet50-2-2-R and ResNet50-2-2-D for [4, 11] km range, shown in Figs.~\ref{fig6}(b) and (c), are representative of the loss for the other models. The corresponding epochs for the training processes are 10 and 2, respectively. From Fig.~\ref{fig6}, the ResNet50-2-x-R requires more training steps than the other models because more classes need to be determined in range estimation. The models with the minimum validation losses are selected as the final models for range and depth prediction.

\subsubsection{\label{subsec:3:4:2} Test data}
It is crucial to examine the localization performance of deep learning models on various test data sets to ensure that the models have generalization capability. In this subsection, only environmental variation is considered (i.e. constant source magnitude and noiseless data). The effects of source magnitudes and SNRs are discussed in Secs.~\ref{subsec:3:4:3} and \ref{subsec:3:4:4}.
\begin{table}[t]
	\caption{\label{tab:table4}Environmental parameters for test data.}	
	\begin{ruledtabular}
		\begin{tabular}{cccccc}
			Parameters & Units & Case1 & Case2 & Case3 & Case4 \\
			\hline
			water depth  & $\rm{m}$ & 69.2 & 72.7 & 70.3 & 71.8 \\
			\hline
			sediment parameters &  &  & & & \\
			sediment thickness & $\rm{m}$ & 8.3 & 19.5 & 12.0 & 23.2 \\
			P-wave speed & $\rm{m/s}$ & 1585 & 1637 & 1615 & 1623 \\
			density & $\rm{g/cm^3}$ & 1.64 & 1.63 & 1.62 & 1.68 \\
			P-wave attenuation & $\rm{dB/\lambda}$ & 0.03 & 0.01 & 0.02 & 0.06 \\
			\hline
			basement parameters &  &  & & & \\
			P-wave speed & $\rm{m/s}$ & 1690 & 1792 & 1725 & 1821 \\
			density & $\rm{g/cm^3}$ & 1.81 & 1.77 & 1.75 & 1.83 \\
			P-wave attenuation & $\rm{dB/\lambda}$ & 0.04 & 0.06 & 0.03 & 0.02 \\
		\end{tabular}
	\end{ruledtabular}	
\end{table}
\begin{figure}[h]	
	\fig{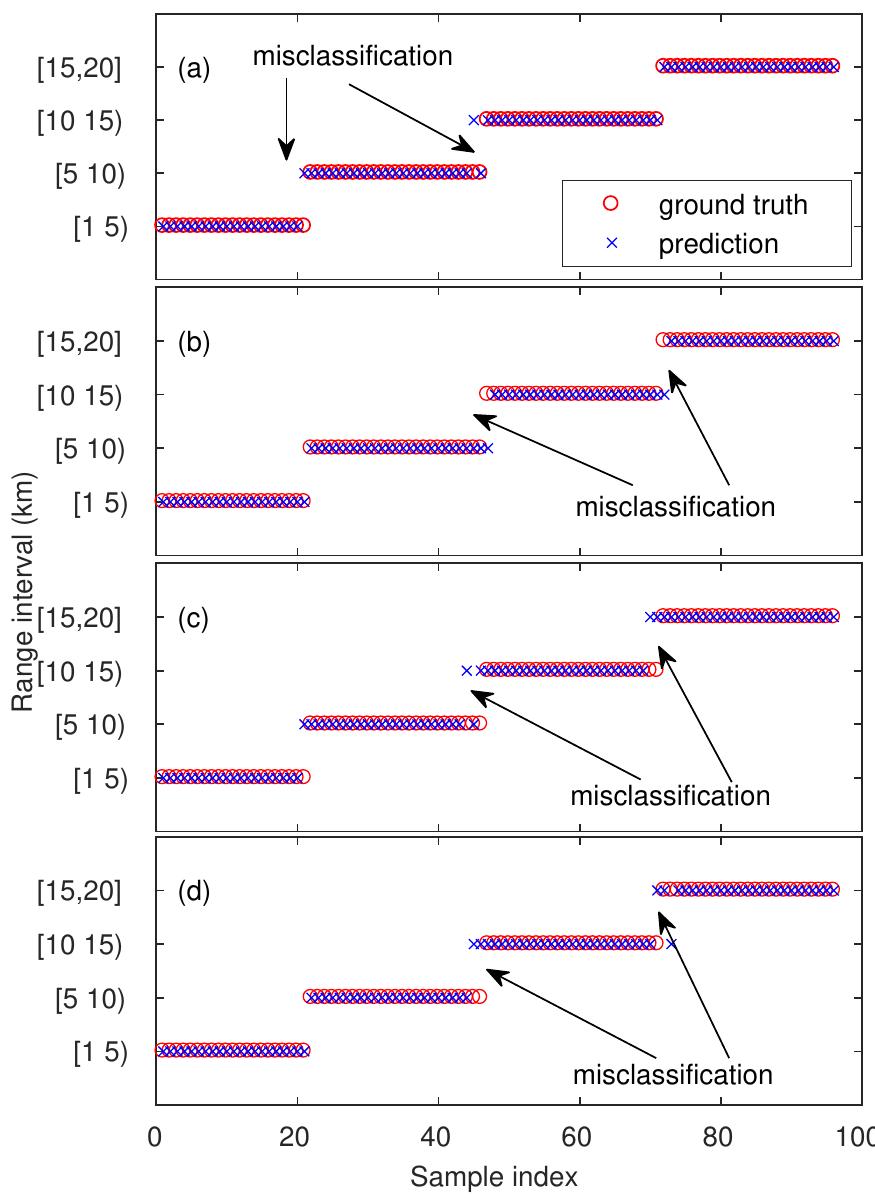}{7.3cm}{}
	\caption{(Color online) Comparison between the ground truth and the predicted range intervals by ResNet50-1 on the environments of (a) Case1, (b) Case2, (c) Case3, and (d) Case4. The samples are generated from 1--20 km with the increment of 0.2 km.}
	\label{fig7}
\end{figure}

\begin{figure*}[tb]		
	\fig{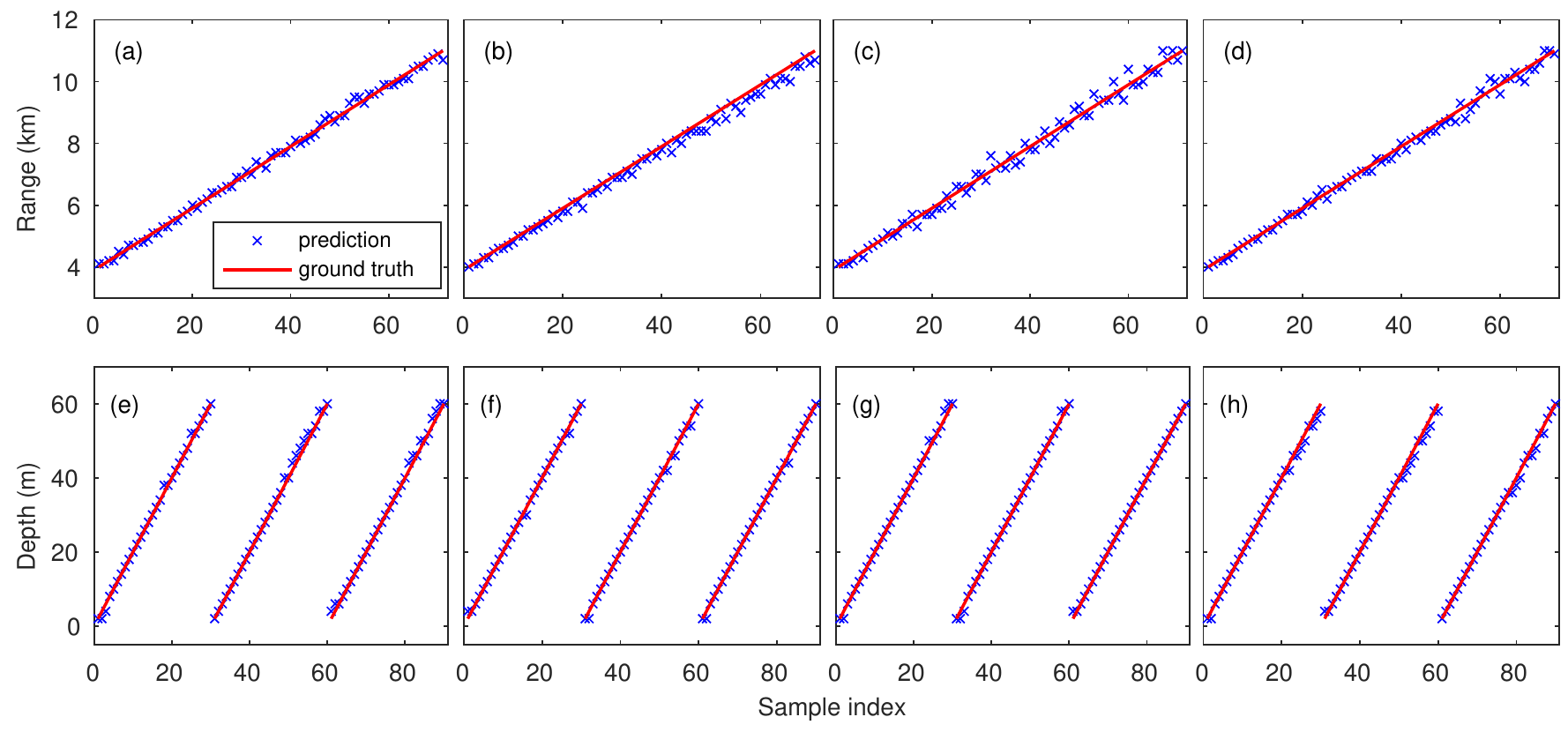}{16.0cm}{}
	\caption{(Color online) Comparison between the ground truth and the predicted source ranges (a)(b)(c)(d) by ResNet50-2-2-R and source depths (e)(f)(g)(h) by ResNet50-2-2-D  on the environments of (a)(e) Case1, (b)(f) Case2, (c)(g) Case3, and (d)(h) Case4. The source depth is fixed at 12 m in (a--d) and three ranges \{6, 8, 10 km\} are used in (e--h).}
	\label{fig8}
\end{figure*}
Four test data sets are generated using the environmental parameters in Table~\ref{tab:table4}. Note that the parameters of cases 1--4 are not part of the parameter set used to generate the training data set, although the water depth and sediment parameters still fall within the bounds. The generated test data sets are used to examine the generalization performance of the ResNet50 models. As in Sec.~\ref{subsec:3:4:1}, the prediction results of ResNet50-1 in the first step and ResNet50-2-2-x for the [4, 11] km in the second step are shown here. 

Figure~\ref{fig7} shows the comparison between the predicted range intervals by the ResNet50-1 and the true values. The samples are generated from the ranges of 1--20 km with the increment of 0.2 km. The source depth used in test set is 12 m. The classification accuracy from Eq.~(\ref{accuracy}) for the cases 1--4 is 97.9, 97.9, 94.8, and 95.8\%, i.e. the number of misclassified samples is 2, 2, 5, and 4 for the cases 1--4. This demonstrates that the ResNet50-1 performs well on different environments. Incorrect classifications occur at the junctions of different range intervals (see Fig.~\ref{fig7}). The design of range overlaps in Table~\ref{tab:table3} for the second step reduces the effect of misclassification, leading to more robust localization.

The predictions for source ranges and depths by the ResNet50-2-2-R and ResNet50-2-2-D are shown in Fig.~\ref{fig8} along with the ground truth. The MAE,  Eq.~(\ref{MAE}), in range predictions for cases 1--4 is 0.08, 0.14, 0.15, and 0.11 km (Figs.~\ref{fig8}(a--d)), while the corresponding error is 0.4, 0.2, 0.2, and 0.5 m in depth estimation (Figs.~\ref{fig8}(e--h)). The ResNet50 models in the second step are able to estimate the source ranges and depths with low errors for different bottom parameters. 

Figures~\ref{fig7} and \ref{fig8} demonstrate that the DNN trained on specific environments can adapt to a variety of environments for noiseless data. It shows the capability of deep learning models in generalization.

\subsubsection{\label{subsec:3:4:3} Source magnitude}

The source magnitude across the frequency spectrum may be not constant. The preprocessing of input data by Eq.~(\ref{normalized_segment}) is used for slowly varying source magnitudes. We use an example to demonstrate it.

\begin{figure*}[t]		
	\fig{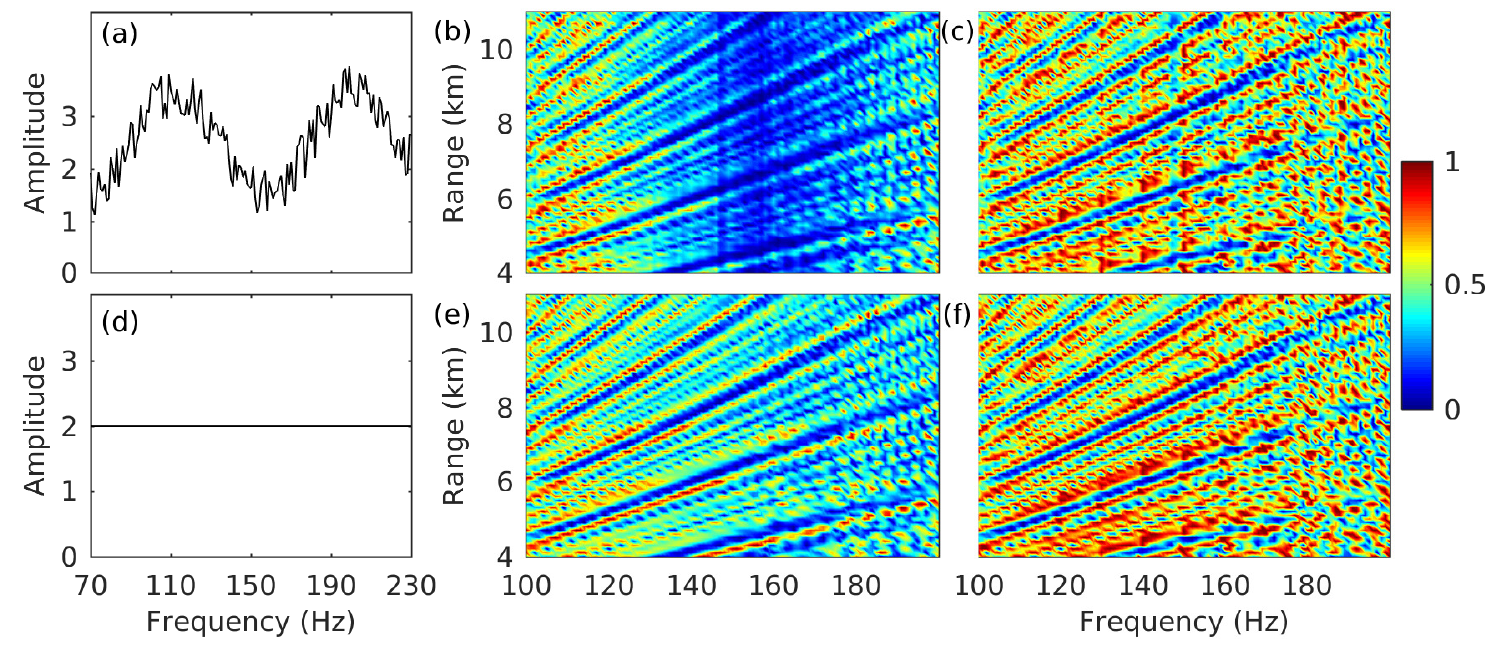}{15.0cm}{}
	\caption{(Color online) Data preprocessing for the source magnitudes slowly varying in frequency domain. One example of slowly varying source magnitude (a) and corresponding test data without (b) and with (c) piecewise normalization of Eq.~(\ref{normalized_segment}). (d)(e)(f) correspond to the same processing for the training data with a flat magnitude. The frequency band 100--200 Hz is used.}
	\label{fig9}
\end{figure*}

Assuming the source magnitude, $|S(f)|=2+\sin{(2\pi f/90)}+rnd$ with $rnd$ denoting the uniformly distributed random number in the interval (0,1), is slowly varying with frequency $f$ as shown in Fig.~\ref{fig9}(a) and the Green's function $g(f,\bf r)$ is calculated by KRAKEN based on the environmental parameters of Case3 in Table~\ref{tab:table4}, the generated test data shown in Fig.~\ref{fig9}(b) are obtained by Eq.~(\ref{normalized_amp}). The source depth and ranges for the test data are 12 m and 4--11 km. The test data with the piecewise normalization by Eq.~(\ref{normalized_segment}) are given in Fig.~\ref{fig9}(c). The results by the same processing Eqs.~(\ref{normalized_amp}) and (\ref{normalized_segment}) for the training data with a flat source magnitude are shown in Figs.~\ref{fig9}(d)(e)(f). As seen in Figs.~\ref{fig9}(a--f), the original frequency spectrogram is transformed by Eq.~(\ref{normalized_segment}) to the one that is independent of the source magnitude. The effect of source magnitude differences between the training and test data is reduced by the use of Eq.~(\ref{normalized_segment}), resulting in more accurate localization.

\subsubsection{\label{subsec:3:4:4} SNR}

In this subsection, the effect of SNR for the test data is investigated. As in Sec.~\ref{subsec:3:4:3}, the bottom parameters of Case3 in Table~\ref{tab:table4} are used for test data generation. The source depths and ranges for the test data are the same as in Sec.~\ref{subsec:3:4:2} for the two steps. Note that the training data are noiseless while the test sets are noisy with different SNRs.
\begin{figure}[h]	
	\fig{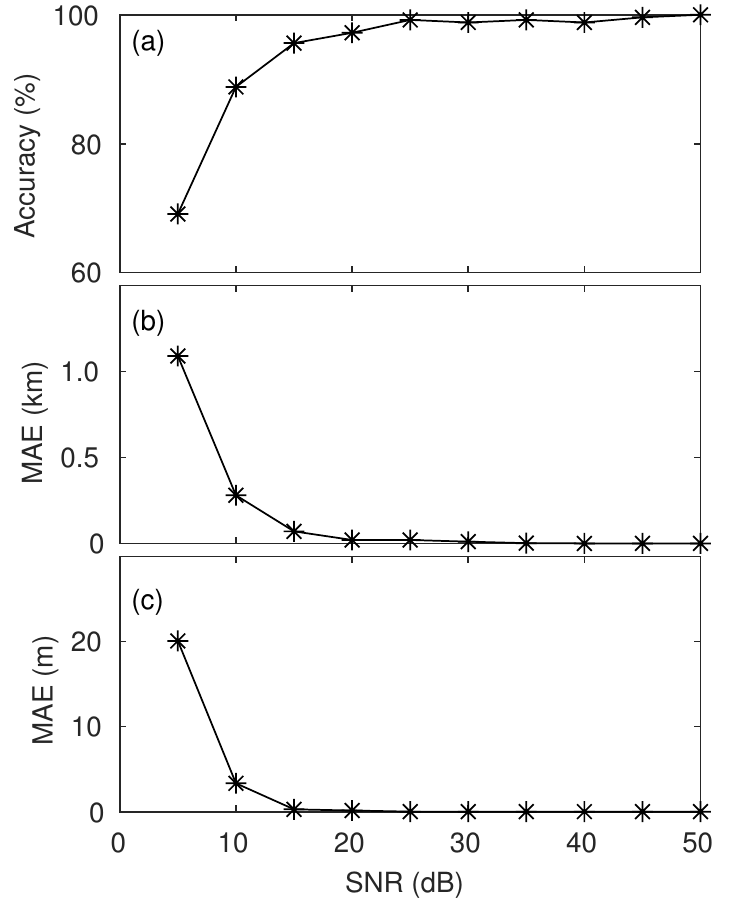}{7.0 cm}{}
	\caption{Performance metrics versus SNR. (a) Prediction accuracy for range interval determination by ResNet50-1; (b) MAE for the range estimation by ResNet50-2-2-R; (c) MAE for the depth estimation by ResNet50-2-2-D.}
	\label{fig10}
\end{figure}

The SNR across the frequency band is defined as
\begin{equation} \label{SNR}
\rm{SNR}=10 \log_{10} \frac{\| \mathbf{p} \|_2^2}{\sigma^2},
\end{equation}
where $\mathbf{p}$ is the complex pressure at $F$ frequencies and $\sigma^2$ represents the noise variance. When adding noise, for each SNR, $\sigma^2$ is calculated separately at each $\mathbf{p}$ computed as a function of range using KRAKEN. 

Figure~\ref{fig10}(a) shows the prediction accuracy calculated by Eq.~(\ref{accuracy}) for different SNRs in the first step. The accuracy improves with SNR and reaches 95.0\% at the SNR of 15 dB. The MAE of range and depth estimation on the test data sets with different SNRs is given in Figs.~\ref{fig10}(b) and (c). As expected, the prediction error for the range and depth reduces from 1.09 km (Fig.~\ref{fig10}(b)) and 20.0 m (Fig.~\ref{fig10}(c)) at the SNR of 5 dB to 0.02 km and 0.13 m at 20 dB. It demonstrates that a relatively high SNR ($\ge$15 dB) is necessary to achieve a convincing prediction. 

\section{Experimental data  \label{sec:4}}

\subsection{\label{subsec:4:1} Experiment description}
\begin{figure}[h]	
	\fig{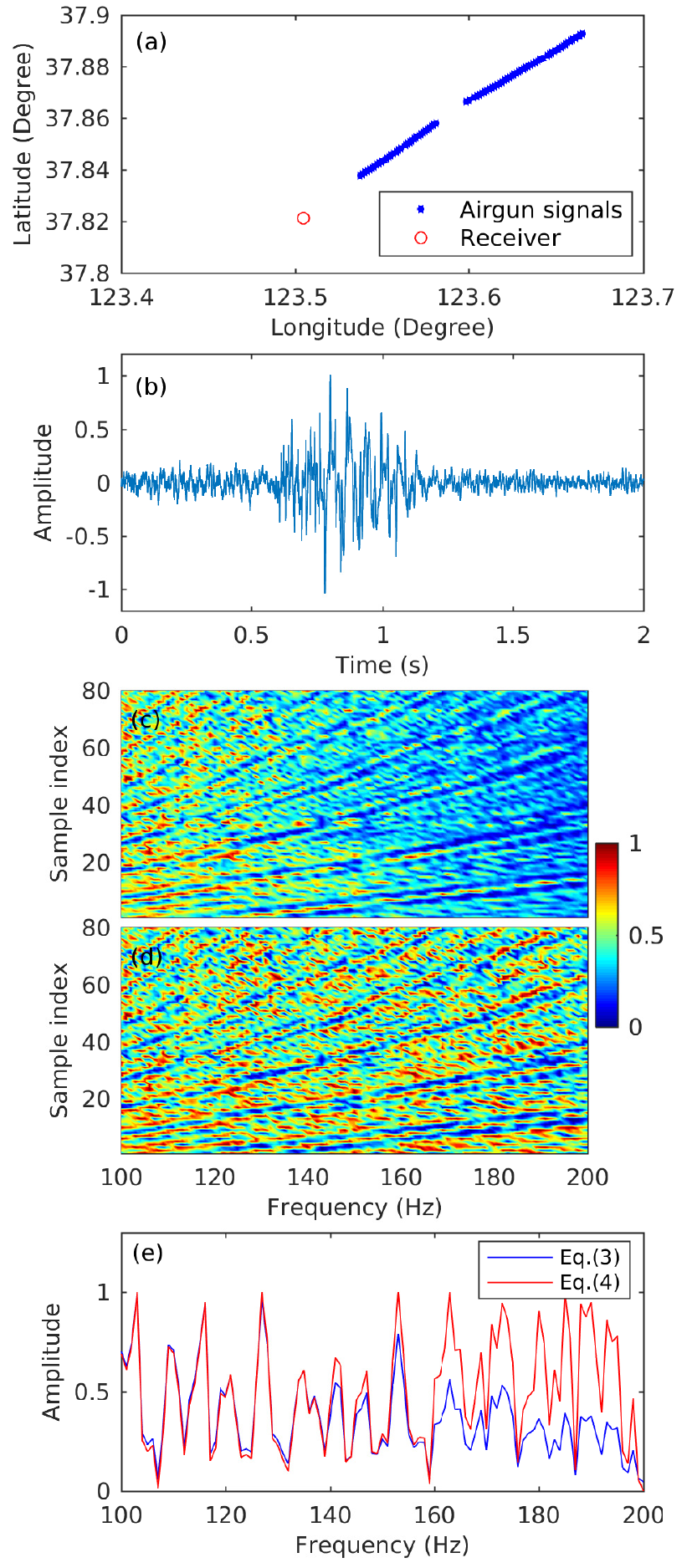}{8.0cm}{}
	\caption{(Color online) (a) Experiment geometry; (b) The waveform of one received airgun signal; Spectrograms of 80 airgun signals normalized using (c) Eq.~(\ref{normalized_amp}) and (d) Eq.~(\ref{normalized_segment}); (e) The spectrum comparison for the 40th signal.}
	\label{fig11}
\end{figure}
The acoustic data were collected during one sea experiment\cite{Ren} conducted in the China Yellow Sea in December 18--20, 2011. The experimental geometry is shown in Fig.~\ref{fig11}(a), with one 'L' shape array (one 16-element vertical array and one 32-element horizontal array) moored on the sea floor. The data recorded by one hydrophone of the horizontal array indicated by the circle in Fig.~\ref{fig11}(a) were used for source localization. The sampling rate for the received signals is 6000 Hz. During the experiment, low-frequency signals were emitted about every one minute in the first 50 minutes of each hour by a towed airgun at ranges 3.0--16.0 km which are denoted by the asterisks in Fig.~\ref{fig11}(a). The ship speed was about 2 m/s. The time-domain waveform of one received airgun pulse is given in Fig.~\ref{fig11}(b). The positions of the airgun are calculated by the GPS data recorded on ship. A depth sensor was attached on the airgun to record the source depth during the experiment. The water depth along the propagation path varies within 68--73 m. In the following data processing, the range-independent assumption is used for the simulated sound field generation. The sound speed profile of water column was approximately constant 1480 m/s as shown in Fig.~\ref{fig4}. 
 
In Fig.~\ref{fig11}(a), there are 80 airgun pulses during a 95-min period corresponding to 3--16 km (47 pulses in the first 50 min, 10 min off, and 33 more pulses in the next 35 min). The duration for each airgun pulse is 2 s. The spectrograms (12000 samples) on one sensor are shown in Figs.~\ref{fig11}(c)(d). Striations existed due to the modal interference. The frequency band used for localization is 100--200 Hz with 1 Hz increment. The spectrum comparison for the 40th signal with Eqs.~(\ref{normalized_amp}) and (\ref{normalized_segment}) is given in Fig.~\ref{fig11}(e), showing that the effect of source magnitude difference across the frequency band is reduced by the piecewise normalization of Eq.~(\ref{normalized_segment}).

\subsection{\label{subsec:4:2} Results}
Source localization is performed following the steps described in Sec.~\ref{subsec:2:3}. As depicted in Fig.~\ref{fig3}, the preprocessed signals are first fed to ResNet50-1 for range interval determination. Subsequently, the inputs are processed by the corresponding models ResNet50-2-x-R and ResNet50-2-x-D for range and depth estimation. The ResNet50 models obtained in Sec.~\ref{sec:3} are tested on the experimental data.

It is of interest to compare the focalized MFP with the deep learning method. The focalized MFP is implemented using SAGA\cite{SAGA} software package (i.e., an implementation based on focalized MFP\cite{Gerstoft1,Siderius} using the environmental parameters in Table~\ref{tab:table2}). The normal mode code SNAP\cite{SAGA} is used to generate the replicas in SAGA. There are 10 unknown parameters including source range and depth searched in SAGA. The optimal parameters are obtained by minimizing the following frequency-coherent objective function:
\begin{equation} \label{MFP}
\phi_F(\Theta)=1-\frac{|\sum_{f=1}^{F} \hat{\mathbf{p}}(f) \hat{\mathbf{q}}(f,\Theta)|^2}{\sum_{f=1}^{F}|\hat{\mathbf{p}}(f)|^2 \sum_{f=1}^{F}|\hat{\mathbf{q}}(f,\Theta)|^2},
\end{equation}
where $\Theta$ represents the unknown parameter set. $\hat{\mathbf{p}}(f)$ and $\hat{\mathbf{q}}(f)$ are the piecewise normalized magnitudes of measured and replica fields (i.e. Eq.~(\ref{normalized_segment})) at the $f$th frequency. 

\begin{figure}[h]	
	\fig{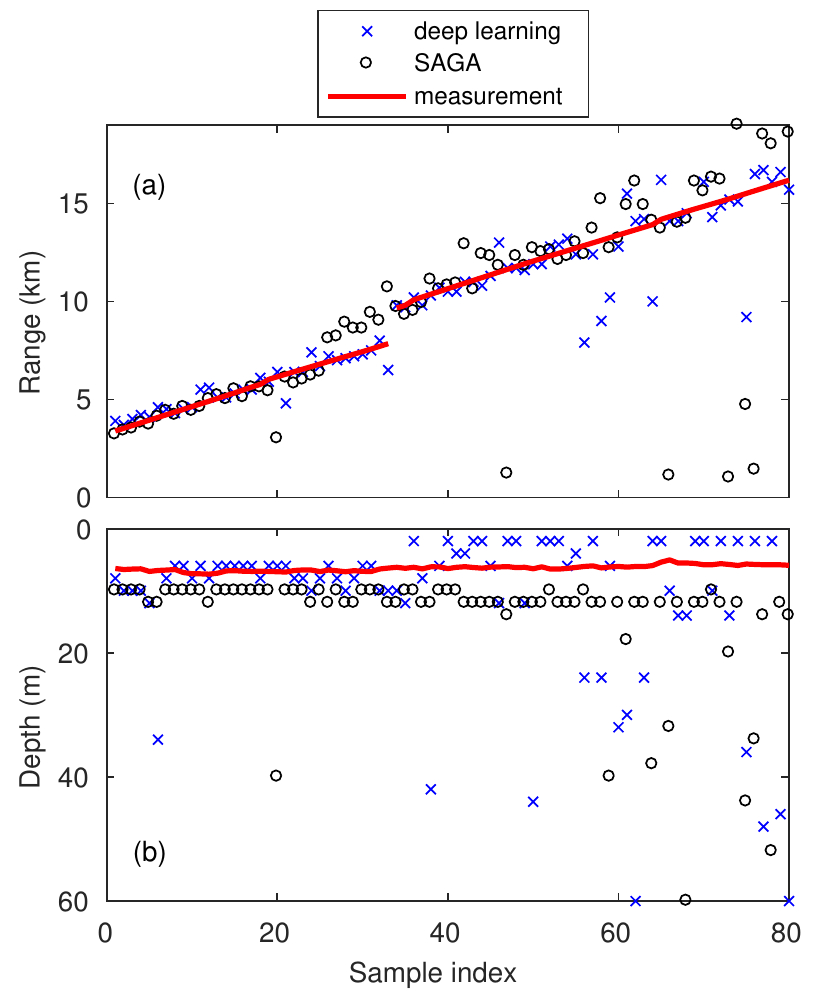}{8.0cm}{}
	\caption{(Color online) Predicted source ranges (a) and depths (b) by the deep learning and SAGA methods.}
	\label{fig12}
\end{figure}
The range and depth estimated by the ResNet50-2-x-R and ResNet50-2-x-D are shown in Figs.~\ref{fig12}(a) and (b), along with the predictions calculated by SAGA. The measurements of ranges and depths are also given in Fig.~\ref{fig12} for comparison. Overall, the range predictions by deep learning and SAGA fit the GPS data well except for some larger errors at distant ranges. The MAE of deep learning calculated by Eq.~(\ref{MAE}) is 0.70 km for range predictions, while it is 1.54 km using SAGA. In terms of the source depths, the airgun was towed at a shallow depth, about 5--8 m according to the measurements. As seen in Fig.~\ref{fig12}(b), most of the depth predictions are within the interval [2, 16] m, agreeing with the experimental configuration. The MAE of depth predictions is 7.7 and 8.1 m for these two methods. The performance for depth estimation degrades at far ranges due to low SNRs. 

\begin{figure}[h]	
	\fig{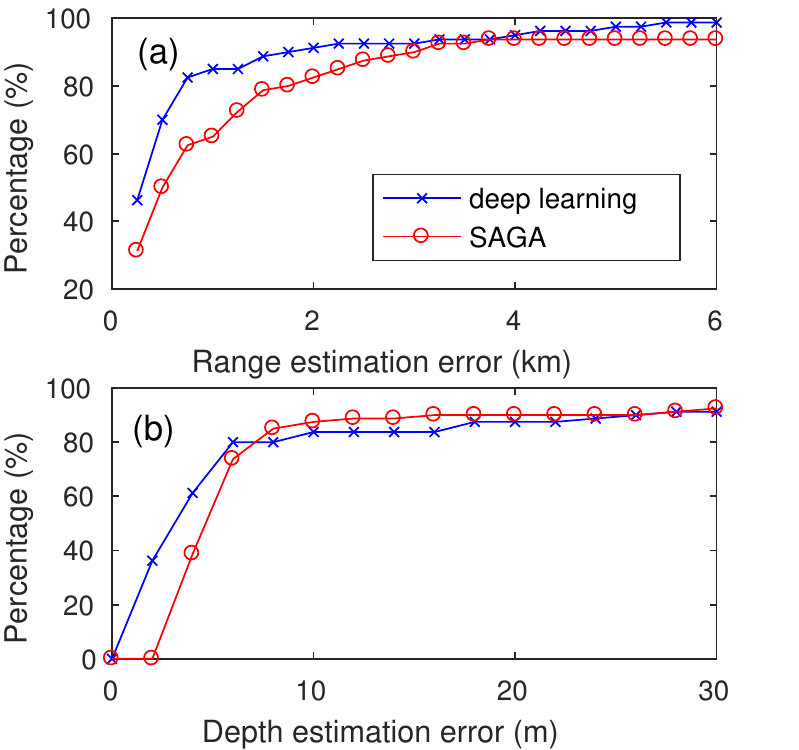}{8.0cm}{}
	\caption{(Color online) Statistic of error percentage for range (a) and depth (b). $y$-axis denotes the proportion of predictions with a maximum error below a specific value, i.e. $x$-axis.}
	\label{fig13}
\end{figure}

The error percentage statistic for range and depth predictions is given in Fig.~\ref{fig13}, which represents the proportion of predictions with a maximum error below a specific value, given on the $x$-axis. The proportion of predictions below 1.5 km error in range is 88.8\% for deep learning and 78.8\% for SAGA, while the proportion of points below 10 m in depth is 83.8 and 87.5\%. It demonstrates that compared with SAGA the performance of the proposed method using deep learning is better in source range estimation and comparable in depth prediction.
 
\section{\label{sec:5} Discussion}

\subsection{\label{subsec:5:1} Computation time}
The localization using DNNs includes the stages of training and prediction. The training stage requires significantly more computation time than prediction, but the models need only be trained once. The training time was 6 days for our ResNet50-1 model and 3 days for each of the ResNet50-2-x-D models. Each ResNet50-2-x-R model took 15 days to train. The models were trained using three NVIDIA TITAN X PASCAL GPU cards. The prediction time for one sample is about 125 milliseconds on one INTEL i7 CPU core (3.4 GHz), demonstrating that real-time localization is achievable with the pre-trained models. For comparison, the prediction time for one sample is about 600 s using SAGA.

\subsection{\label{subsec:5:2} Extension to low-SNR and multi-source cases }
As stated in Sec.~\ref{subsec:3:4:4}, the proposed localization approach is only convincing for signals with high SNRs since only noiseless data are used in the training process. Similarly, the approach is only valid for a single source localization. However, this method can be used for low-SNR and multi-source (with different directions) cases with a horizontal array using conventional array signal processing. In the first step, the direction-of-arrival (DOA) estimation for multiple sources is performed by beamforming techniques. In the second step, for each source, the beamformed signal at the source direction (i.e. spatial filtering) with a higher SNR is obtained. The enhanced signal after spatial filtering is then fed into the deep learning models for localization as described in the paper.

\subsection{\label{subsec:5:3} Limitations and future work}
This paper deals with the source localization with uncertain bottom properties. There are some limitations for the application of our algorithm:

(1) Time-varying SSPs of water column (e.g. in summer) were not considered. Therefore, the effect of varying SSPs on localization needs further study.

(2) The water depth was restricted to a small interval to reduce the size of training data set. Thus, the DNNs should be re-trained for significantly different ocean environments (e.g. the water depth).  

(3) The range-independent propagation environments were hypothesized to generate the training data. The localization for range-dependent cases needs further investigation.

\section{\label{sec:6} Summary}

This paper presents an approach for source localization using deep residual neural networks based on synthetic data which are generated by an acoustic propagation model. In our algorithm, absolute pressure on only one sensor is used for source range and depth estimation. A two-step training strategy is proposed to alleviate the training difficulty for the deep models with big data. The range intervals are first determined and then the source ranges and depths are solved by the deep learning models corresponding to the specific range intervals.

The performance of ResNet50 models were tested on simulated test data sets with different bottom parameters, source magnitudes and SNRs. The results show that the proposed ResNet50 models perform well on various environments, slowly varying source magnitudes, and high SNRs. The experimental data further verified the localization performance of our approach in uncertain environments, where 88.8\% range predictions have the absolute error below 1.5 km and the error of 83.8\% depth estimates is within 10 m.

\begin{acknowledgments}
This research was supported by the National Natural Science Foundation of China (Grant Nos. 11434012 and 11874061) and the Youth Innovation Promotion Association, Chinese Academy of Sciences.
\end{acknowledgments}





\begin{thebibliography}{99}

\bibitem{Niu1} H. Niu, E. Reeves, and P. Gerstoft, ``Source localization in an ocean waveguide using supervised machine learning,'' J. Acoust. Soc. Am. \textbf{142}, 1176--1188, (2017).

\bibitem{Niu2} H. Niu, E. Ozanich, and P. Gerstoft, ``Ship localization in Santa Barbara Channel using machine learning classifiers,'' J. Acoust. Soc. Am. \textbf{142}, EL455--EL460, (2017).

\bibitem{Bucker} H. P. Bucker, ``Use of calculated sound fields and matched field detection to locate sound source in shallow water,'' J. Acoust. Soc. Am. \textbf{59}, 368--373, (1976).

\bibitem{Westwood}  E. K. Westwood, ``Broadband matched-field source localization,'' J. Acoust. Soc. Am. \textbf{91}, 2777--2789, (1992).

\bibitem{Baggeroer} A. B. Baggeroer, W. A. Kuperman, and P. N. Mikhalevsky, ``An overview of matched field methods in ocean acoustics,'' IEEE J. Ocean. Eng. \textbf{18}, 401--424, (1993).

\bibitem{Michalopoulou} Z. H. Michalopoulou and M. B. Porter, ``Matched-field processing for broadband source localization,'' IEEE J. Ocean. Eng. \textbf{21}, 384--392, (1996).
		
\bibitem{Steinberg}  B. Z. Steinberg, M. J. Beran, S. H. Chin and J. H. Howard, ``A neural network approach to source localization,'' J. Acoust. Soc. Am. \textbf{90}, 2081--2090, (1991).

\bibitem{Ozard}  J. M. Ozard, P. Zakarauskas and P. Ko, ``An artificial neural network for range and depth discrimination in matched field processing,'' J. Acoust. Soc. Am. \textbf{90}, 2658--2663, (1991).

\bibitem{Michalopoulou2} Z. H. Michalopoulou, D. Alexandrou, and C. Moustier, ``Application of neural and statistical classifiers to the problem of seafloor characterization,'' IEEE J. Ocean. Eng. \textbf{20}, 190--197, (1995).

\bibitem{Lefort} R. Lefort, G. Real, and A. Dr\'emeau, ``Direct regressions for underwater acoustic source localization in fluctuating oceans,'' Appl. Acoust.  \textbf{116}, 303--310, (2017).		
	
\bibitem{Wang} Y. Wang and H. Peng, ``Underwater acoustic source localization using generalized regression neural network,'' J. Acoust. Soc. Am. \textbf{143}, 2321--2331, (2018).

\bibitem{Huang} Z. Huang, J. Xu, Z. Gong, H. Wang, and Y. Y, ``Source localization using deep neural networks in a shallow water environment,'' J. Acoust. Soc. Am. \textbf{143}, 2922--2932, (2018).

\bibitem{Liu} Y. Liu, H. Niu, and Z. Li, ``Source ranging using ensemble convolutional networks in the direct zone of deep water,'' Chin. Phys. Lett. \textbf{36}, 044302-1--044302-4, (2019).

\bibitem{Gemba1} K. L. Gemba, W. S. Hodgkiss, and P. Gerstoft, ``Adaptive and compressive matched field processing,'' J. Acoust. Soc. Am. \textbf{141}, 92--103, (2017).

\bibitem{Gemba2} K. L. Gemba, S. Nannuru, P. Gerstoft, and W. S. Hodgkiss, ``Multi-frequency sparse Bayesian learning for robust matched field processing,'' J. Acoust. Soc. Am. \textbf{141}, 3411--3420, (2017).

\bibitem{Gerstoft1} P. Gerstoft, ``Inversion of seismoacoustic data using genetic algorithms and a posteriori probability distributions,'' J. Acoust. Soc. Am. \textbf{95}, 770--782, (1994).

\bibitem{Gingras} D. F. Gingras and P. Gerstoft, ``Inversion for geometric and geoacoustic parameters in shallow water: Experimental results,'' J. Acoust. Soc. Am. \textbf{97}, 3589--3598, (1995).

\bibitem{Collins} M. D. Collins and W. A. Kuperman, ``Focalization: Environmental focusing and source localization,'' J. Acoust. Soc. Am. \textbf{90}, 1410--1422, (1991).

\bibitem{Dosso1} S. E. Dosso, ``Matched-field inversion for source localization with uncertain bathymetry,'' J. Acoust. Soc. Am. \textbf{94}, 1160--1163, (1993).

\bibitem{Baer} R. N. Baer and M. D. Collins, ``Source localization in the presence of gross sediment uncertainties,'' J. Acoust. Soc. Am. \textbf{120}, 870--874, (2006).

\bibitem{Richardson} A. M. Richardson and L. W. Nolte, ``A posteriori probability source localization in an uncertain sound speed, deep ocean,'' J. Acoust. Soc. Am. \textbf{89}, 2280--2284, (1991).

\bibitem{Dosso2} S. E. Dosso and M. J. Wilmut, ``Uncertainty estimation in simultaneous Bayesian tracking and environmental inversion,'' J. Acoust. Soc. Am. \textbf{124}, 82--97, (2008).

\bibitem{Dosso3} S. E. Dosso and M. J. Wilmut, ``Comparison of focalization and marginalization for Bayesian tracking in an uncertain ocean environment,'' J. Acoust. Soc. Am. \textbf{125}, 717--722, (2009).

\bibitem{Frazer} L. N. Frazer and P. I. Pecholcs, ``Single-hydrophone localization,'' J. Acoust. Soc. Am. \textbf{88}, 995--1002, (1990).

\bibitem{Jesus} S. M. Jesus, M. B. Porter, Y. St\'ephan, X. D\'emoulin, O. C. Rodr\'iguez, and E. M. Coelho, ``Single hydrophone source localization,'' IEEE J. Ocean. Eng. \textbf{25}, 337--346, (2000).

\bibitem{Siderius} M. Siderius, P. Gerstoft, and P. Nielsen, ``Broadband geoacoustic inversion from sparse data using genetic algorithms,'' J. Comput. Acoust. \textbf{6}, 117--134, (1998).

\bibitem{Hermand} J. P. Hermand, ``Broad-band geoacoustic inversion in shallow water from waveguide impulse response measurements on a single hydrophone: theory and experimental results,'' IEEE J. Ocean. Eng. \textbf{24}, 41--66, (1999).

\bibitem{Gac} J. C. Le Gac, M. Asch, Y. St\'ephan, and X. Demoulin, ``Geoacoustic inversion of broad-band acoustic data in shallow water on a single hydrophone,'' IEEE J. Ocean. Eng. \textbf{28}, 479--493, (2003).

\bibitem{Spain} G. L. D'Spain and W. A. Kuperman, ``Application of waveguide invariants to analysis of spectrograms from shallow water environments that vary in range and azimuth,'' J. Acoust. Soc. Am. \textbf{106}, 2454--2468, (1999).

\bibitem{Cockrell} K. L. Cockrell and H. Schmidt, ``Robust passive range estimation using the waveguide invariant,'' J. Acoust. Soc. Am. \textbf{127}, 2780--2789, (2010).

\bibitem{Rakotonarivo} S. T. Rakotonarivo and W. A. Kuperman, ``Model-independent range localization of a moving source in shallow water,'' J. Acoust. Soc. Am. \textbf{132}, 2218--2223, (2012).

\bibitem{Deep1} G. Hinton and R. R. Salakhutdinov, ``Reducing the dimensionality of data with neural networks," Science \textbf{313}, 504--507. (2006).

\bibitem{Deep2}  Y. LeCun, Y. Bengio, and G. Hinton, ``Deep learning," Nature \textbf{521}, 436--444, (2015).

\bibitem{Deep3}  J. Schmidhuber, ``Deep learning in neural networks: An overview," Neural Networks \textbf{61}, 85--117, (2015).

\bibitem{Krizhevsky} A. Krizhevsky, I. Sutskever, and G. E. Hinton, ``Imagenet classification with deep convolutional neural networks," Adv. Neural Inf. Process. Syst., 1097--1105, (2012).

\bibitem{Hinton} G. Hinton, L. Deng, D. Yu, G. E. Dahl, A. Mohamed, N. Jaitly, A. Senior, V. Vanhoucke, P. Nguyen, T. N. Sainath, and B. Kingsbury, ``Deep neural networks for acoustic modeling in speech recognition: The shared views of four research groups," IEEE Signal Proc. Mag. \textbf{29}, 82--97, (2012).

\bibitem{Collobert} R. Collobert, J. Weston, L. Bottou, M. Karlen, K. Kavukcuoglu, and P. Kuksa,  ``Natural language processing (almost) from scratch," J. Mach. Learn. Res. \textbf{12}, 2493--2537, (2011).

\bibitem{Porter} M. B. Porter, The KRAKEN Normal Mode Program, http://oalib.hlsresearch.com/Modes/AcousticsToolbox/manualtml/
kraken.html (Last viewed 11/1/2009).

\bibitem{He} K. He, X. Zhang, S. Ren, and J. Sun, ``Deep residual learning for image recognition," In \textit{Proceedings of the IEEE conference on computer vision and pattern recognition}, 770--778, (2016).

\bibitem{cs231n} \url{http://cs231n.github.io/convolutional-networks/\#overview} (Last viewed January 14, 2019).

\bibitem{Chollet} F. Chollet, ``Keras: Deep learning library for theano and tensorflow," \textit{URL: https://keras.io}, (2015).

\bibitem{Tensorflow} M. Abadi, A. Agarwal, P. Barham, E. Brevdo, Z. Chen, C.Citro, G. S. Corrado, A. Davis, J. Dean, M. Devin, S. Ghemawat, I. Goodfellow, A. Harp, G.Irving, M. Isard, Y. Jia, R. Jozefowicz, L. Kaiser, M. Kudlur, J. Levenberg, D. Man\'e, R. Monga, S. Moore, D. Murray, C. Olah, M. Schuster, J. Shlens, B. Steiner, I. Sutskever, K. Talwar, P. Tucker, V. Vanhoucke, V. Vasudevan, F. Vi\'egas, O. Vinyals, P. Warden, M. Wattenberg, M. Wicke, Y. Yu, and X. Zheng, ``TensorFlow: Large-scale machine learning on heterogeneous distributed systems," \emph{Software available from tensorflow.org}, (2015).

\bibitem{Ren} Y. Ren and Y. Qi,  ``Waveguide invariant and range estimation based on phase-shift-compensation of underwater acoustic spectrograms," AIP Conf. Proc. \textbf{1495}, 627-633, (2012).

\bibitem{SAGA} P. Gerstoft, SAGA users guide: an inversion software package, \url{http://noiselab.ucsd.edu/saga/saga.html} (Last viewed January 16, 2019).

	

	
	
	
\end{thebibliography}

\end{document}